\documentclass[prl,twocolumn]{revtex4-2}

\usepackage{graphicx}
\usepackage{dcolumn}
\usepackage{bm}
\usepackage{amsmath}

\begin{document}


\title{Universal Scaling and Many-Body Resurrection of Polaritonic Double-Quantum Coherences}

\author{Maxim Sukharev}
\email{maxim.sukharev@asu.edu}
\affiliation{Department of Physics, Arizona State University, Tempe, Arizona 85287, United States}
\affiliation{College of Integrative Sciences and Arts, Arizona State University, Mesa, Arizona 85212, United States} 

\date{\today}

\begin{abstract}
The ultrafast nonlinear optical response of molecular ensembles is fundamentally altered under strong light-matter coupling. To rigorously isolate the genuine many-body contributions, an exact time-domain field-subtraction protocol is developed within a fully non-perturbative Maxwell-Liouville framework explicitly incorporating the two-exciton manifold in real space and time. This approach reveals that while collective cavity delocalization drives the macroscopic nonlinear signal toward a severe harmonic cancellation (an effect termed "spectral starvation"), intrinsic many-body molecular interactions robustly resurrect genuine polaritonic double-quantum coherences (DQCs). This many-body resurrection is governed by a universal two-photon matching rule, $\Delta_B + 4J = \Omega_R$, linking molecular anharmonicity ($\Delta_B$) to the macroscopic Rabi splitting ($\Omega_R$) and excitonic coupling ($J$). Crucially, this resonance exploits the spatial mismatch between macroscopic polaritons and localized two-exciton pairs to break harmonic cancellation. For J-aggregates ($J < 0$), this condition uniquely isolates the resonant many-body state below the dense manifold of localized dark states, protecting the macroscopic coherence from spatial fragmentation. This predictive framework establishes a direct phase diagram to engineer and protect optical nonlinearities across diverse strongly coupled platforms.
\end{abstract}

\maketitle

Molecular polaritons, hybrid quasiparticles arising from the strong coupling between confined optical modes and molecular transitions, offer an unprecedented avenue to fundamentally control photochemistry, energy transport, and optical nonlinearities \cite{Ebbesen2016HybridPerspective, Ribeiro2018PolaritonCavities, Dunkelberger2022Vibration-CavityDynamics, Fregoni2022TheoreticalChemistry, Simpkins2023ControlCoupling, Bhuyan2023ThePhotophysics, McKillop2025APolaritons, Biswas2025EmergentField, Basov2025PolaritonicMatter}. To interrogate the complex many-body dynamics of these systems, ultrafast nonlinear techniques, such as two-dimensional (2D) and pump-probe spectroscopies, have emerged as vital experimental tools \cite{Xiang2018Two-dimensionalPolaritons, Xiang2020IntermolecularCoupling, Delpo2020PolaritonCoupling, Mewes2020EnergySpectroscopy, Chen2022Cavity-enabledPseudorotation, Duan2021IsolatingSpectra, Son2022EnergySpectroscopy, Pyles2024RevisitingModes, Sufrin2024ProbingSpectroscopyc, Yin2025OvercomingPolaritons, Sufrin2026Phase-ResolvedMeta-Surfaces,Debnath2025CoherentSpectroscopy}. However, interpreting the ultrafast nonlinear optical response of molecular ensembles in microcavities presents a profound theoretical challenge and requires moving beyond traditional approximations to account for genuine many-body correlations \cite{F.Ribeiro2018TheoryPolaritons, Sidler2021PolaritonicProperties, Ribeiro2021EnhancedCoupling, Sidler2022ACollectivity, Fowler-Wright2022EfficientPolaritons, Mandal2022TheoreticalElectrodynamics, Zhang2023MultidimensionalApproach, Bauman2025PerspectiveSystems}. 

The core complication arises from collective delocalization. In a perfectly harmonic system, the macroscopic third-order nonlinear response strictly vanishes due to exact destructive interference between competing excitation pathways \cite{Spano1989NonlinearSize, Spano1991CooperativeBehavior, Mukamel1999PrinciplesSpectroscopy,Debnath2026PhotonAggregates}. Resonant optical cavities force strongly coupled molecular ensembles toward this harmonic limit, severely suppressing genuine polaritonic double-quantum coherences (DQCs). This effect, termed spectral starvation, effectively buries the essential microscopic many-body correlations. Standard theoretical models relying on mean-field approximations, single-exciton truncations, or the rotating-wave approximation inherently fail to capture the subtle anharmonicities, two-particle interactions, and retardation effects required to resurrect and resolve these signals \cite{Ribeiro2021EnhancedCoupling, Reitz2025NonlinearDynamics, Pyles2024RevisitingModes}.

Capturing these many-body signatures requires moving beyond the standard Tavis-Cummings paradigm \cite{Abramavicius2009CoherentPerspectives}. Conventional models treat molecular ensembles as independent two-level systems homogeneously coupled to a single cavity mode \cite{Perez-Sanchez2023SimulatingModels}. While applicable to cavity-confined monomers \cite{Debnath2022EntangledDephasing}, this approximation fundamentally fails in the macroscopic many-body regime. It entirely omits the near-field excitonic couplings and intrinsic biexciton binding energies that dictate microscopic correlations. Furthermore, because the macroscopic cavity field continuously re-excites the extended molecular ensemble, standard perturbative expansions of the density matrix in powers of the external field break down. Cavity-mediated back-action inextricably mixes linear and higher-order polarizations, rendering standard diagrammatic Liouville pathways inadequate for strongly coupled systems.

To overcome these theoretical limitations, this Letter introduces an exact time-domain field-subtraction protocol developed within a fully non-perturbative Maxwell-Liouville framework. Within this self-consistent framework, the molecular system is rigorously described by an effective Hamiltonian $\hat{H} = \hat{H}_0 + \hat{H}_{\mathrm{int}}(t)$ \cite{Spano1991BiexcitonMonolayers, Abramavicius2009CoherentPerspectives, Gutierrez-Meza2026FrenkelAggregates}. The static part, $\hat{H}_0$, accounts for the single- and two-exciton manifolds in the site basis. The light-matter interaction is governed by $\hat{H}_{\mathrm{int}}(t)$, where $\Omega_i(t)$ is the local Rabi frequency: 

\begin{widetext}
\begin{equation}
\hat{H}_{0} = \sum_{i} \omega_{e} |e_{i}\rangle\langle e_{i}| + \sum_{i \neq j} J_{ij} |e_{i}\rangle\langle e_{j}| + \sum_{i < j} (2\omega_{e} - \Delta_{\mathrm{B}}) |f_{ij}\rangle\langle f_{ij}|,
\end{equation}
\begin{equation}
\hat{H}_{\mathrm{int}}(t) = -\sum_{i}\Omega_{i}(t)(|e_{i}\rangle\langle g|+\text{h.c.}) -\sum_{i<j} (\Omega_{j}(t)|f_{ij}\rangle\langle e_{i}| + \Omega_{i}(t)|f_{ij}\rangle\langle e_{j}| + \text{h.c.}).
\end{equation}
\end{widetext}
Here, employing units of $\hbar=1$, $\omega_{e}$ is the bare molecular resonance, $J_{ij}$ represents the inter-site excitonic coupling, and $\Delta_{\mathrm{B}}$ defines the intrinsic molecular anharmonicity (or biexciton binding energy) that rigidly shifts the two-exciton manifold spanned by $|f_{ij}\rangle$. The two-exciton states are treated in a localized-pair basis; inter-pair hopping is omitted, as justified in the Supplemental Material for hard-core bosons in the regime $\Omega_R>4|J_{ij}|$, where $\Omega_R$ is the macroscopic Rabi splitting.
\begin{figure}[t]
    \centering
    \includegraphics[width=\columnwidth]{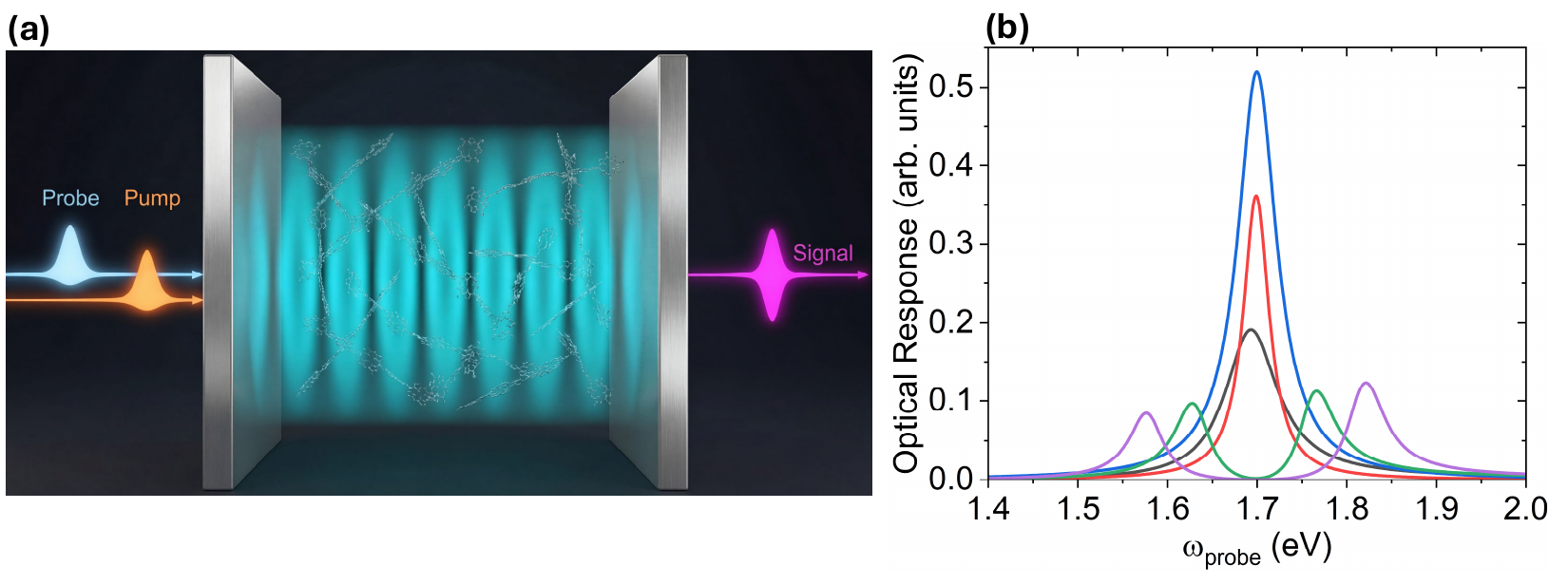}
    \caption{(a) Schematic of the pump-probe homodyne detection setup for a molecular ensemble strongly coupled to an optical cavity. (b) Linear optical response establishing the strong coupling regime. The black line shows transmission for the empty cavity. Red and blue lines display the out-of-cavity molecular absorption for low ($3\times10^{19}$ cm$^{-3}$, red) and high ($10^{20}$ cm$^{-3}$, blue) molecular densities. The corresponding in-cavity transmission profiles (green and purple lines) illustrate the macroscopic Rabi splitting and the formation of the lower and upper polariton ($\omega_\mathrm{LP}$ and $\omega_\mathrm{UP}$) branches. Mirror-to-mirror distance is 305.5 nm.}
    \label{fig1}
\end{figure}

To capture the full electrodynamics of the light-matter interaction, the time evolution of the spatially dependent density matrix is evaluated. The quantum master equation explicitly incorporates independent channels for non-radiative decay and pure dephasing at each spatially distributed site [see Supplemental Material]. This quantum system is coupled self-consistently to macroscopic Maxwell's equations. Such a semiclassical Maxwell-Liouville framework captures exact propagation and retardation effects in real space and time without invoking the rotating-wave approximation.

As illustrated in Fig. \ref{fig1}(a), the physical system consists of an optical cavity formed by two gold mirrors. A thin molecular layer is positioned at the cavity center to ensure maximum light-matter interaction. The separation between the mirrors is tuned to align the fundamental cavity mode with the excitonic transition $\omega_e$. Specific geometric dimensions and material parameters are detailed in the Supplemental Material.

\begin{figure}[t]
    \centering
    \includegraphics[width=\columnwidth]{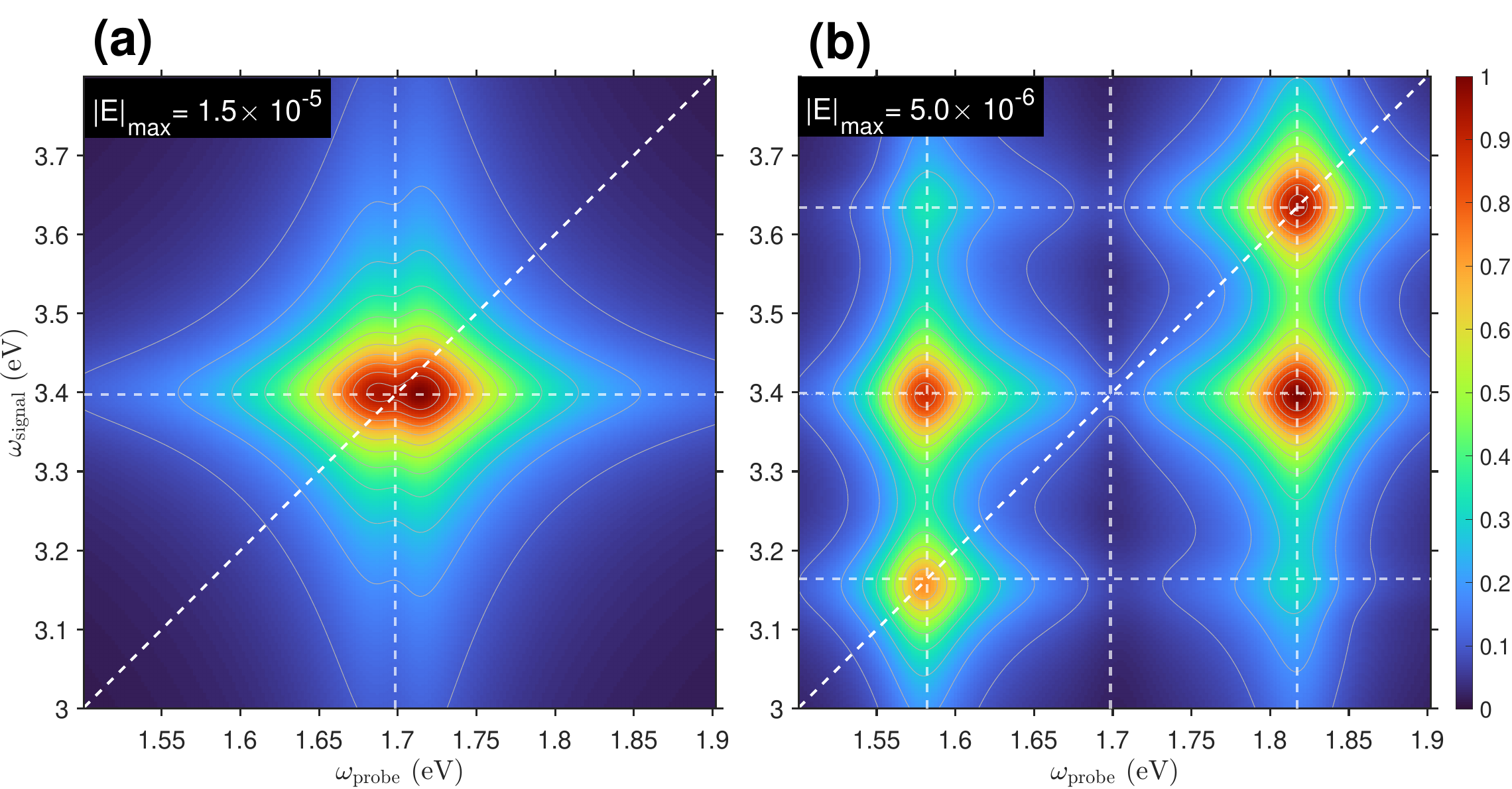}
    \caption{Breakdown of the harmonic approximation and the onset of spectral starvation. 2D homodyne signal amplitude for the high-density ($10^{20}$ cm$^{-3}$) system evaluated at zero molecular anharmonicity ($\Delta_B=0$). (a) Out-of-cavity (slab) response exhibiting a single peak at the bare molecular resonances. (b) In-cavity response showing polaritonic splitting. The white dashed grid indicates the fundamental ($\omega_\mathrm{LP}$, $\omega_\mathrm{UP}$) and harmonic sum frequencies ($2\omega_\mathrm{LP}$, $2\omega_\mathrm{UP}$, $\omega_\mathrm{LP}+\omega_\mathrm{UP}$). Despite identical molecular densities, the maximum field amplitude in the cavity is starved relative to the bare slab, reflecting the suppression of population-modulation artifacts by the collective delocalization of the single-exciton manifold.}
    \label{fig2}
\end{figure}

The spectroscopy protocol utilizes a broadband pump pulse to create a coherent population within the polaritonic manifold, which is subsequently queried by a time-delayed probe pulse. This 1-pump/1-probe setup serves as an ideal laboratory for probing DQC migration and many-body interactions in the strong coupling regime. Prior to nonlinear excitation, the linear response shown in Fig. \ref{fig1}(b) establishes the baseline. With the bare cavity tuned to $\omega_e$, the formation of distinct lower and upper polariton ($\omega_\mathrm{LP}$ and $\omega_\mathrm{UP}$) branches is observed. Two density regimes are compared to investigate how collective delocalization dictates the fate of the subsequent nonlinear response.

\begin{figure*}[t]
    \centering
    \includegraphics[width=\textwidth]{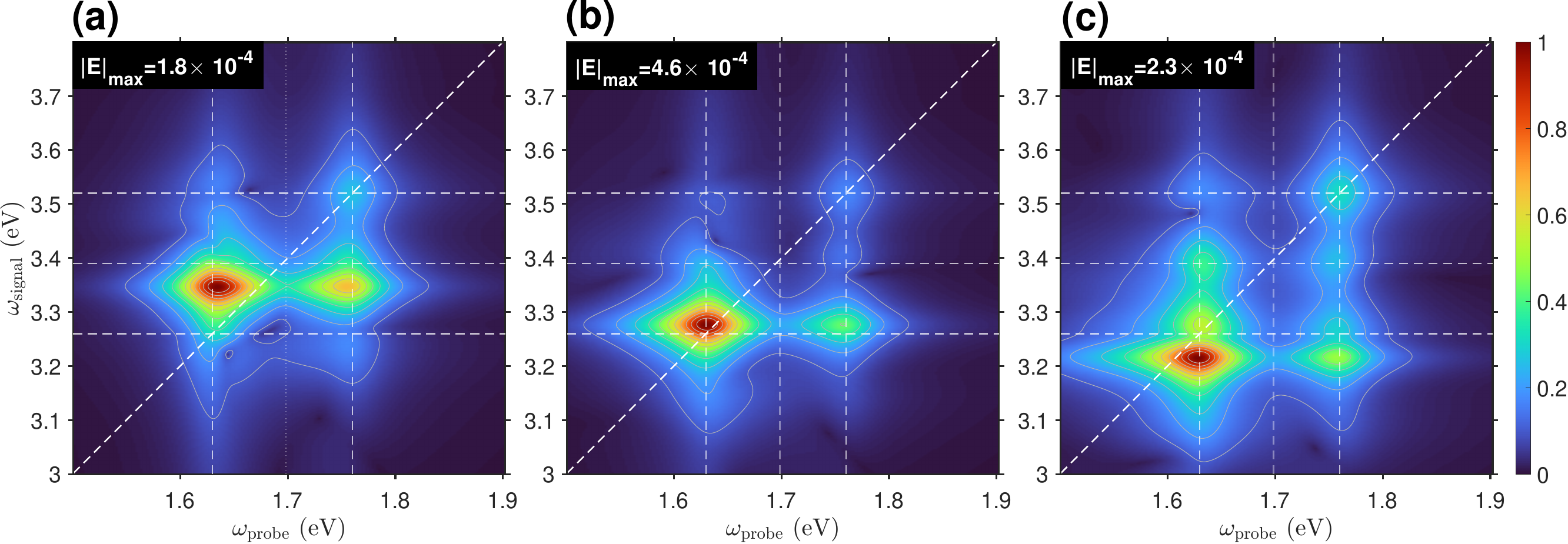}
    \vspace{-20pt}
    \caption{Spectral migration and many-body resurrection of the genuine double-quantum coherence (DQC). The in-cavity low-density response ($3\times10^{19}$ cm$^{-3}$) is shown for increasing molecular anharmonicities: (a) $\Delta_B = 50$ meV, (b) $\Delta_B = 120$ meV, and (c) $\Delta_B = 180$ meV. The genuine DQC peak systematically detaches from the harmonic sum frequencies of the single-exciton manifold (indicated by the dashed grid). Maximum resurrection is observed in panel (b), where the molecular anharmonicity, $\Delta_B$, matches the Rabi frequency, $\Omega_R$, maximizing the coupling between the polariton states and the localized many-body manifold. The white dashed lines crossing in the center of each panel denote the uncoupled bare molecule resonances ($\omega_e$ and $2\omega_e$) for reference.}
    \label{fig3}
\end{figure*}

To isolate elusive DQC signatures, an exact time-domain field-subtraction protocol is implemented:
\begin{equation}
\mathbf{E}_{\mathrm{NL}}(t) = \mathbf{E}_{\mathrm{pump+probe}}(t) - \mathbf{E}_{\mathrm{pump}}(t) - \mathbf{E}_{\mathrm{probe}}(t).
\end{equation}
Here, $\mathbf{E}_{\mathrm{pump+probe}}(t)$ is the total transmitted field when both pulses interact with the system simultaneously, while $\mathbf{E}_{\mathrm{pump}}(t)$ and $\mathbf{E}_{\mathrm{probe}}(t)$ are the individual transmitted fields when each pulse propagates through the cavity independently. This approach strictly isolates the nonlinear interaction field, which is the phase-stable signal arising from the mutual interaction of the pulses. The semiclassical Maxwell-Liouville framework remains formally non-perturbative and inherently accounts for saturation effects and cavity-mediated back-action. Executing the subtraction at the field level preserves the full interferometric phase stability required for homodyne-detected 2D spectroscopy while circumventing the prohibitive numerical overhead of conventional phase-cycling. By Fourier transforming $\mathbf{E}_{\mathrm{NL}}(t)$ with respect to the pump-probe delay and the detection time, the 2D spectra are constructed [see Supplemental Material for signal processing details].

The isolated nonlinear interaction field is analyzed in the frequency domain to extract DQC signatures. The time-domain subtraction protocol is specifically designed to isolate the $E_{\mathrm{pump}}^2 E_{\mathrm{probe}}$ pathway. Because the broadband pump interacts twice without a temporal delay, conventional single-quantum coherence pathways collapse to $\omega_{\mathrm{signal}} = 0$. This rigorously isolates the pure DQC-like region where the signal frequency $\omega_{\mathrm{signal}}$ strictly corresponds to twice the probe energy. Fig. \ref{fig2}(a) displays the resulting 2D homodyne signal for the out-of-cavity high-density molecular slab. The spectrum exhibits a prominent splitting of the maximum along the $\omega_{\mathrm{probe}}$ axis. This feature is a direct physical manifestation of the optical density of the molecular layer, an effect inherently resolved by propagating Maxwell's equations directly in real space and time.

Fig. \ref{fig2}(b) demonstrates the corresponding in-cavity response evaluated in the perfectly harmonic limit ($\Delta_B = 0$). The spectrum features distinct peaks at the fundamental and sum frequencies of the lower and upper polaritons. Notably, these peaks do not represent genuine two-photon emission. Instead, they are expected population-modulation artifacts arising from the nonlinear driving of the system. This is confirmed by their persistence in isolated single-pump simulations restricted purely to a single-exciton model. While a genuine many-body DQC depends strongly on the anharmonicity $\Delta_B$, it remains completely buried by these sum-frequency artifacts in the harmonic limit. Furthermore, a direct comparison of the absolute field amplitudes between the bare slab and the coupled cavity reveals a profound attenuation of the macroscopic nonlinear signal. This phenomenon is termed ``spectral starvation''. It reflects the suppression of population-modulation artifacts dictated by the collective delocalization of the single-exciton manifold.

The suppression of the macroscopic nonlinear signal observed in the harmonic limit is dramatically reversed upon the introduction of molecular anharmonicity, as illustrated in Fig. \ref{fig3}. As the molecular anharmonicity $\Delta_B$ is systematically increased, the genuine many-body DQC detaches from the stationary background of population-modulation artifacts and migrates continuously along the $\omega_{\mathrm{signal}}$ axis. The residual sum-frequency peaks remain energetically pinned throughout this evolution, firmly establishing their origin within the single-exciton manifold and their complete independence from the anharmonic shift. 

Remarkably, this spectral migration is accompanied by a massive enhancement of the nonlinear response. Near the optimal resonance condition [Fig. \ref{fig3}(b)], the peak field amplitude exceeds that of the starved harmonic cavity by approximately two orders of magnitude, constituting a genuine many-body resurrection of the macroscopic signal. It is imperative to note that resolving this profound recovery strictly necessitates the explicit integration of the two-exciton manifold. Conventional mean-field approximations or single-exciton truncations inherently fail to capture the physics of this localized many-body state, underscoring the indispensable nature of the fully correlated semiclassical model utilized here.

\begin{figure*}[t]
    \centering
    \includegraphics[width=\textwidth]{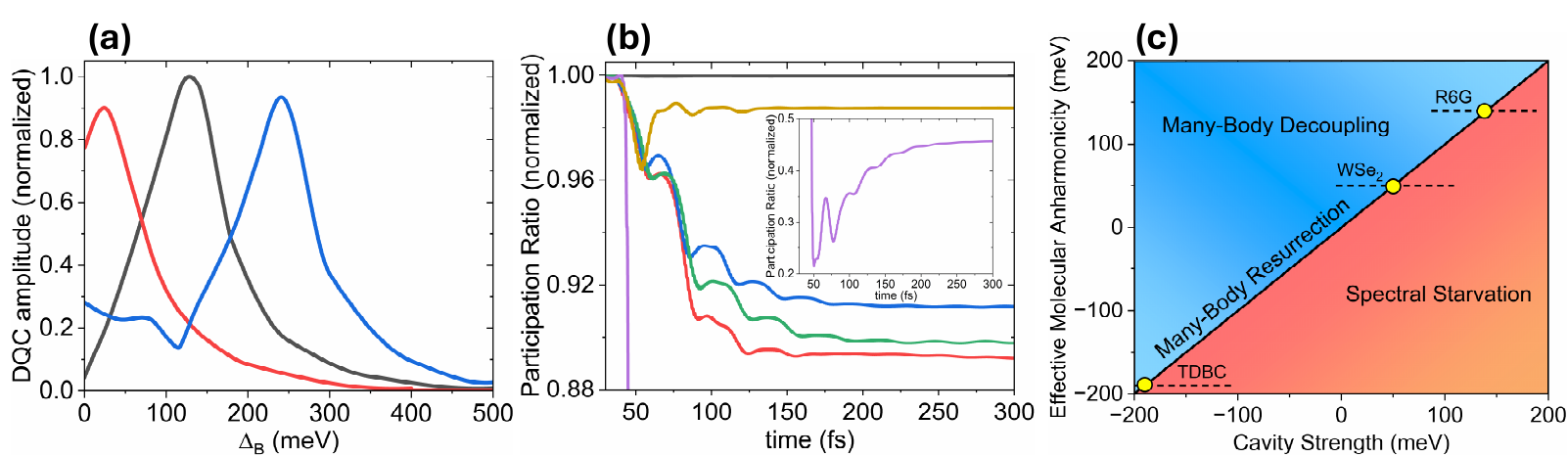}
    \caption{Scaling of the nonlinear signal amplitude, time-resolved spatial delocalization, and the universal phase diagram. For each $J$-case, the bare cavity is tuned to the maximum molecular absorption of the respective aggregate. (a) In-cavity peak DQC amplitudes for the low-density system comparing the uncoupled molecular control (black; $J=0$ meV) with H-aggregate (red; $J=30$ meV) and $J$-aggregate (blue; $J=-30$ meV) configurations. All curves are normalized to the peak signal of the $J=0$ control. The many-body interactions within the molecular manifold fundamentally rewire the cavity-mediated suppression or enhancement of the nonlinear response. (b) Time-resolved participation ratio normalized to the total number of distinct site pairs ($\binom{N}{2} = 351$). Curves correspond to: the uncoupled control (black; $\Delta_B = 128$ meV); H-aggregates (red, blue; $\Delta_B = 0, 23$ meV); and J-aggregates at the baseline (green; $\Delta_B = 0$), the localization trap (purple; $\Delta_B = 120$ meV), and the resurrected resonance (gold; $\Delta_B = 240$ meV). The inset highlights the rapid spatial collapse within the localization trap at $\Delta_B = 4|J|$. The protection plateau (gold) demonstrates the recovery of collective delocalization, tracking the $J=0$ baseline. (c) Universal phase diagram for the many-body response. The boundary $\Delta_B + 4J = \Omega_R$ defines the "Many-Body Resurrection" where intrinsic molecular interactions balance cavity-induced delocalization. Regions of Spectral Starvation (harmonic regime) and Many-Body decoupling (localized regime) are indicated. Representative materials (TDBC, WSe$_2$, and R6G) are mapped along their respective tuning trajectories (dashed lines) with yellow markers identifying the optimal resonance points.}
    \label{fig4}
\end{figure*}

The parametric dependencies governing the macroscopic scaling of the nonlinear response are fundamentally mapped by the strict two-photon double resonance condition presented in Fig. \ref{fig4}(a). Analyzing the amplitude maxima establishes a universal spectroscopic matching rule for the resurrection of the many-body signal in a strongly coupled system:
\begin{equation}
\label{Delta_law}
\Delta_{\mathrm{B}} + 4J = \Omega_{\mathrm{R}}.
\end{equation}
This condition dictates that maximum DQC amplitude is uniquely achieved when the molecular anharmonicity, $\Delta_B$, perfectly balances the macroscopic Rabi splitting, $\Omega_R$, and excitonic coupling, $J$. At this matching point, the energetic ladder is perfectly detuned, maximizing the bimodal signal by optimizing the sequential two-photon transition through the protected lower polariton (LP) state to the localized biexciton manifold ($2\omega_{\mathrm{LP}} = 2\omega_e-\Delta_B$). This universal scaling seamlessly transitions the DQC maximum from $\Delta_B = \Omega_R$ for uncoupled monomers ($J=0$) to significantly higher binding energies for J-aggregates ($J<0$), where the bright state resides at the bottom of the single-exciton band.

To rigorously quantify the spatial extent of the DQC, the normalized participation ratio, $P$, with respect to the total number of sites is evaluated \cite{Thouless1974ElectronsLocalization, Scholes2020LimitsAggregates}. This metric tracks the number of molecular pairs actively contributing to the DQC and is defined based on bimodal off-diagonal density matrix elements $P_{i,j} = |\rho_{f_{ij},g}|^2$:
\begin{equation}
P = \frac{(\sum_{i<j} P_{i,j})^2}{\sum_{i<j} P_{i,j}^2}.
\end{equation}
A normalized participation ratio approaching unity indicates a DQC perfectly shared across the entire macroscopic ensemble, while sub-unity values signify spatial localization.

The dynamics of the participation ratio in Fig. \ref{fig4}(b) confirms that both uncoupled monomers (black line) and optimally resonant J-aggregates (gold line) sustain near-perfect delocalization ($P \approx 1$). However, a catastrophic collapse in spatial coherence is observed for J-aggregates at exactly $\Delta_{\mathrm{B}} = 4|J|$, corresponding to a local minimum of the DQC signal seen in Fig. \ref{fig4}(a), blue line. Under this condition, the participation ratio drops to very small values [Fig. \ref{fig4}(b), purple line]. This specific anharmonicity defines a purely molecular localization trap independent of the cavity. At this point, the bound $f$-states satisfy accidental harmonic spacing with the bottom of the bare molecular band. This harmonic cancellation forces the giant, mobile bright mode to go completely silent, leaving only highly fragmented and localized dark states to sustain a minimal residual signal \cite{Spano1991BiexcitonMonolayers}. H-aggregates ($J>0$) similarly struggle to maintain delocalization [Fig. \ref{fig4}(b), red and blue lines]. Because their matching requires a lower $\Delta_B$, the target many-body state overlaps with the dense molecular scattering continuum, causing rapid bimodal dephasing and spatial fragmentation.

These competing microscopic mechanisms culminate in the definitive universal phase diagram for polaritonic nonlinearities presented in Fig. \ref{fig4}(c). The boundary rigidly defined by Eq. (\ref{Delta_law}) separates distinct regimes of nonlinear operation: collective harmonic spectral starvation and the nonresonant many-body decoupling observed at excessively high anharmonicities. This predictive mapping provides a precise architectural blueprint for engineering robust nonlinearities in strong-coupling platforms. It establishes that J-aggregated dyes, which aggressively isolate localized many-body states below the scattering continuum, are uniquely protected to maximize macroscopic optical nonlinearities in optical cavities.

In conclusion, the exact semiclassical electrodynamic framework presented here resolves the fundamental competition between collective harmonicity and microscopic many-body interactions in strongly coupled molecular ensembles. While resonant optical cavities inherently suppress macroscopic nonlinearities through spectral starvation, intrinsic molecular anharmonicity provides a robust mechanism for signal resurrection. This recovery is strictly governed by a universal two-photon matching rule that optimizes the coherent transition through the protected lower polariton to the localized biexciton manifold. These findings establish a direct, predictive design paradigm for polaritonic chemistry and ultrafast quantum technologies. By actively engineering the interplay between macroscopic cavity detuning and microscopic excitonic coupling, robust pure many-body optical nonlinearities can be systematically optimized and protected.

Financial support from the Office of Naval Research (Grant No. N000142512090) and the Air Force Office of Scientific Research (Grant No. FA9550-25-1-009) is gratefully acknowledged. The author also expresses sincere gratitude to Prof. Abraham Nitzan and Prof. Lev Chuntonov for numerous invaluable and insightful discussions.

\bibliography{references}
\end{document}